\documentstyle[preprint,aps]{revtex}
\title{Hermitian boson mapping and finite truncation}
\author{Calvin W.~Johnson and Joseph N.~Ginocchio}
\address{Theoretical Division, Los Alamos National Laboratory, Los Alamos, NM
87545}

\begin{document}
\draft

\maketitle
\begin{abstract}
\noindent
Starting from a general, microscopic fermion-to-boson mapping that
preserves Hermitian conjugation, we discuss truncations of the boson
Fock space basis.  We give conditions under which the exact boson images
of finite fermion operators are also finite (e.g., a 1+2-body fermion
Hamiltonian is mapped to a 1+2-body boson Hamiltonian) in the
truncated basis.  For the most general case, where the image is not
necessarily exactly finite, we discuss how to make practical and
controlled approximations.
\end{abstract}


Pairwise correlations are often important in describing the physics of
many-fermion systems.
The classic paradigm is the BCS theory of superconductivity\cite{BCS},
where the wavefunction is dominated
by Cooper pairs which have electrons coupled up to zero linear momentum and
spin; these boson-like pairs condense into a coherent wavefunction.
Another example is the phenomenological Interacting Boson
Model (IBM) for nuclei\cite{IBM},
where many states and transition amplitudes are
successfully described using only $s$- and $d$- (angular momentum
$J=0,2$) bosons,
which represent coherent nucleon pairs.  In both cases
the large number of fermion degrees of freedom are well modelled by only a
few boson degrees of freedom.  On the other hand, however, despite some
forays by Otsuka et al.\ \cite{OAI}, a rigorous microscopic basis for such
phenomenological models is lacking.

The basic problem is to represent the underlying fermion dynamics and
statistics with a boson image amenable to approximation and numerical
calculation. Considerable effort has gone into mapping fermion pairs into
bosons \cite{Klein,Ring:Schuck}.  However
 these mappings typically suffer from a variety of defects.  Most, such as
the Belyaev-Zelevinskii \cite{BZ} and Marumori \cite{Maru} mappings give
rise to boson Hamiltonians with infinite expansions, that is, $k$-body
terms with $k \rightarrow \infty$.
Convergence is slow even
when ``collective'' fermion pairs are used. Finite  but non-Hermitian boson
Hamiltonians have also been derived, such as the well-known Dyson
mapping \cite{Dyson,Rowe}.  Such non-Hermitian mappings can mix
`physical' and non-physical or `spurious' states \cite{doba} as
discussed below.

Less well-known are finite Hermitian alternatives to the Dyson mapping,
obtained by mapping commutation relations of pairs of fermion creation
and annihilation operators \cite{kuchta,marshalek}.  These alternatives
also have drawbacks: they either require a complicated, infinite operator
(our ``norm'' operator defined below) to project out the correct
Fermi statistics, or else the entire boson Fock space must be retained,
which is far more unwieldy than the original fermion problem and which
defeats the purpose of boson mappings.

In this paper, by an alternate approach that maps matrix elements rather
than commutation relations, we find that matrix elements of finite
(specifically 1- or 2-body) fermion operators can be represented, in the
full Fock space, by a finite boson image times an infinite boson ``norm''
or projection operator; because the image does not mix physical and spurious
states projection is not needed. Furthermore the image of a Hermitian fermion
operator is also Hermitian. This regains the result contained in
\cite{marshalek}. We then give conditions under which one can obtain
a finite and exact image in a truncated boson Fock space, and illustrate with
an example.   In the most general case the exact image is not necessarily
finite, but we show how to construct higher-order $k$-body
terms in a systematic
and practical fashion which could lead to effective approximations.

Consider a fermion space with $2\Omega$ single-particle states, and
a fermion Hamiltonian $\hat{H}$. The general problem is to solve the
fermion eigenvalue equation
\begin{equation}
\hat{H}\left | \Psi_p \right \rangle =
E_p \left | \Psi_p \right \rangle,
\label{eigeneqn}
\end{equation}
find transition amplitudes between eigenstates, and so on.  To do this we
require a many-body basis.
Often the basis set for many-fermion wavefunctions are Slater determinants,
antisymmeterized products of single-fermion wavefunctions which we can write
as  products of the Fock creation operators $a^{\dagger}_j, j = 1, \cdot \cdot
\cdot, 2 \Omega$ on the vacuum $a^{\dagger}_{i_{1}} \cdot \cdot \cdot
a^{\dagger}_{i_{n}} \left | 0 \right \rangle $ for $n$ fermions.  These states
span the antisymmetric
irreducible representation of the unitary group in
$2 \Omega$ dimensions, ${\rm SU}(2\Omega)$.
But for an even number of fermions one can just as well
construct states from $N = n/2$ fermion pair creation operators,
\begin{equation}
 \left | \psi_\beta \right \rangle
= \stackrel{N}{\prod_{m=1}} \hat{A}^\dagger_{\beta_m} \left | 0 \right
\rangle,
\label{WfnDefn}
\end{equation}
with
\begin{equation}
\hat{A}^\dagger_\beta \equiv  { 1 \over \sqrt{2}} \sum_{ij}
\left ( {\bf A}^\dagger_\beta \right )_{ij}
{a}^\dagger_i {a}^\dagger_j.
\nonumber
\end{equation}
We always choose the  $\Omega(2\Omega-1)$ matrices ${\bf A}_\beta$ to be
antisymmetric to
preserve the underlying fermion statistics, and we choose for the
normalization the trace $ {\rm tr \,}  \left ( {\bf A}_\alpha {\bf
A}^\dagger_\beta \right )
= \delta_{\alpha \beta}$.
For this paper we  represent generic one- and two-body operators by
\begin{equation}
\hat{T} \equiv
\sum_{ij} T_{ij} {a}^\dagger_i {a}_j, \, \, \, \,
\hat{V} \equiv \sum_{\mu \nu} \left \langle \mu
\right | V \left |  \nu \right \rangle
\hat{A}^\dagger_\mu \hat{A}_\nu.
\label{DefV}
\end{equation}

We begin with the straightforward mapping to boson states
\begin{equation}
 \left | \psi_\beta \right \rangle
\rightarrow  \left | \phi_\beta \right )
= \stackrel{N}{\prod_{m=1}} {b}^\dagger_{\beta_m} \left | 0 \right ),
\label{BosonMap}
\end{equation}
where the ${b}^\dagger$ are boson creation operators.  In conjunction
with this simple mapping of states we construct boson {\it representations}
 that follow
the philosophy of the Marumori expansion
\cite{Maru} and preserve matrix elements of the fermion operators, for
example introducing boson representations $\hat{\cal T}_B$, $\hat{\cal V}_B$,
and most importantly $\hat{\cal H}_B$ such that
$ \left ( \phi_\alpha \right | \hat{\cal T}_B \left | \phi_\beta \right )
= \left \langle \psi_\alpha \right| \hat{T} \left | \psi_\beta \right
\rangle$,
$
\left ( \phi_\alpha \right | \hat{\cal V}_B \left | \phi_\beta \right )
= \left \langle \psi_\alpha \right| \hat{V} \left | \psi_\beta \right \rangle,
$
and
$ \left ( \phi_\alpha \right | \hat{\cal H}_B \left | \phi_\beta \right )
= \left \langle \psi_\alpha \right| \hat{H} \left | \psi_\beta \right
\rangle$.
In addition, because wavefunctions of the form (\ref{WfnDefn}) do not form
an orthonormal set,  we construct the norm operator $\hat{\cal N}_B$ with
the property
$ \left ( \phi_\alpha \right | \hat{\cal N}_B \left | \phi_\beta \right )
= \left \langle \psi_\alpha \right. \left | \psi_\beta \right \rangle$.
The construction of these operators, which will be given in detail in
\cite{JG},
follow directly from the matrix elements which are found by generalizing
the vector coherent state method of Reference \cite{RoweVarPair}.
This procedure differs from the the usual Marumori expansion in that the
latter does not have an explicit norm operator.
With this mapping the fermion eigenvalue equation (\ref{eigeneqn}) becomes
a generalized (because of the norm) boson eigvenvalue equation
\begin{equation}
\hat{\cal H}_B \left | \Phi_p \right ) =
E_p   \hat{\cal N}_B \left | \Phi_p \right ).
\label{BosonEigen1}
\end{equation}
Because we have defined our boson operators so as to preserve matrix elements,
the original energy spectrum of (\ref{eigeneqn}) is found in
(\ref{BosonEigen1}).  However, because the boson space is much larger than
the original fermion space, (\ref{BosonEigen1}) also has additional
spurious states that do not correspond to physical fermion states.  These
by construction have zero eigenvalue and do not mix with the physical space.

When one constructs the norm operator \cite{JG} one finds it can be
conveniently expressed in terms of the  $k$th order Casimir operators of
the unitary group ${\rm SU}(2\Omega)$, $\hat{C}_k = 2^k \,{\rm tr \,} \left(
{\bf P }
\right)^k$, ${\bf P }=
\sum_{\sigma \tau}
{b}^\dagger_\sigma {b}_\tau {\bf A}_\sigma {\bf A}^\dagger_\tau$ (and so is
both a matrix and a boson operator; the trace is over the matrix indices
and not the boson Fock space)
\begin{equation}
\hat{\cal N}_B
= \, \colon \exp \left ( -
\sum_{k=2}^\infty
{ (-1)^{k} \over 2k}
\hat{C}_k \right ) \colon
\label{NormCasimirRep}
\end{equation}
where the colons `:' refer to normal-ordering of the boson operators.
This result is also found in \cite{doba}.
Expanding (\ref{NormCasimirRep}) in normal order one obtains
the form \cite{JG}
\begin{equation}
\hat{\cal N}_B
= 1 + \sum_{k = 2}^\infty
\sum_{ \sigma \tau}
w_k^0 \left ( \left \{ \sigma \right \} ;
\left \{ \tau \right \} \right)
\prod_{m = 1}^k {b}^\dagger_{\sigma_m}
\prod_{n = 1}^k {b}_{\tau_n}.
\label{NormBosonRep}
\end{equation}
The $k$-body boson terms embody the fact that
fermion pair creation and annihilation operators do not have the
same commutation relations as do boson operators, and act to enforce the
Pauli principle.

Although
the representations $\hat{\cal T}_B, \hat{\cal V}_B$ are also complicated
many-body
operators similar in form to (\ref{NormBosonRep}),
we can write them in compact form,
a result we have not seen previously in the literature,
\begin{eqnarray}
\hat{\cal T}_B
 = 2 \sum_{\sigma,\tau} \colon {\rm tr \, }
\left[ {\bf A}_\sigma {\bf T} {\bf A}^\dagger_\tau {\bf G} \right]
b^\dagger_\sigma b_\tau \hat{ \cal N}_B \colon ,
\label{FullOneBody} \\
\hat{\cal V}_B =
\sum_{\mu, \nu}
\left \langle \mu \left | V \right | \nu \right \rangle
\sum_{\sigma, \tau} \colon \left \{
{\rm tr \,} \left [ {\bf A}_\sigma {\bf A}^\dagger_\mu {\bf G} \right ]
{\rm tr \,} \left [ {\bf A}_\nu {\bf A}^\dagger_\tau {\bf G} \right ]
\right. \nonumber \\
\left.
+ 4 \,{\rm tr \,} \left [
{\bf A}_\sigma {\bf A}^\dagger_\mu {\bf P G A}_\nu {\bf A}^\dagger_\tau
{\bf G} \right ] \right \}
b^\dagger_\sigma b_\tau  \hat{ \cal N}_B \colon ,
\label{FullTwoBody}
\end{eqnarray}
where ${\bf G} = ({\bf 1} + 2{ \bf P})^{-1}$
\cite{JG}.
In general the boson representations given in (\ref{NormCasimirRep}),
(\ref{FullOneBody}) and (\ref{FullTwoBody})
do not have good convergence properties,
so that simple termination of the series in $k$-body terms as in, e.g.,
(\ref{NormBosonRep})  is impossible and use of the generalized eigenvalue
equation
(\ref{BosonEigen1}), as written, is problematic.

The explicit forms of
(\ref{FullOneBody}), (\ref{FullTwoBody})
 suggests, however, that these representations factor in a
simple way:
$\hat{\cal T}_B = \hat{\cal N}_B \hat{T}_B =  \hat{T}_B \hat{\cal N}_B$ and
$\hat{\cal V}_B = \hat{\cal N}_B \hat{V}_B =  \hat{V}_B \hat{\cal N}_B$,
where the factored operators $\hat{T}_B, \, \hat{V}_B$, which we term  the
boson images of  $\hat{T}, \, \hat{V}$, have
simple forms.  For example,  a one-body fermion operator has a one-body boson
image
\begin{equation}
 \hat{T}_B =
2\sum_{\sigma \tau}
{\rm tr \,}
\left ({\bf A}_\sigma {\bf T A}_{\tau}^\dagger
 \right)
{b}^\dagger_{\sigma} {b}_\tau.
\end{equation}
To prove factorization, one puts $\hat{T}_B \hat{\cal N}_B$ into
normal order:
\begin{equation}
2\sum_{\sigma \tau} {\rm tr} \left ({\bf A}_\sigma {\bf T A}_{\tau}^\dagger
 \right)
\left \{
{b}^\dagger_{\sigma} \hat{\cal N}_B
{b}_\tau
- {b}^\dagger_{\sigma} \left [ \hat{\cal N}_B,
{b}_\tau \right ] \right \} ,
\end{equation}
and uses \cite{JG}   the completeness relation
\begin{equation}
\sum_\alpha
\left( {\bf A}^\dagger_\alpha \right )_{ij}
\left( {\bf A}_\alpha \right )_{j^\prime i^\prime}
= {1 \over 2} \left( \delta_{i,i^\prime}\delta_{j,j^\prime}
- \delta_{i,j^\prime}\delta_{j,i^\prime}\right ).
\label{completeness}
\end{equation}
and the resulting identities
\begin{equation}
2 \sum_{\alpha} {\rm tr} ({\bf Q A}^{\dagger}_{\alpha}) {\rm tr} ({\bf
A}_{\alpha} {\bf R}) = {\rm tr} ({\bf Q R}) - {\rm tr} ({\bf Q}^T
  {\bf R}),
\end{equation}
\begin{equation}
2 \sum_{\alpha} {\rm tr} ({\bf Q A}^{\dagger}_{\alpha}
 {\bf R} {\bf  A}_{\alpha}) = {\rm tr} ({\bf Q ) }{\rm tr} ({\bf R ) }
 - {\rm tr} ({\bf Q}^T
  {\bf R}).
\end{equation}
We can also show that $[\hat{T}_B, \hat{\cal N}_B] = 0$ from these identities.

The two-body interaction $V$ defined in (\ref{DefV})
can be rewritten in terms of products of two one-body  fermion
operators plus a remainder one-body by rearranging the
fermion Fock operators.
We can then map these one-body operators in terms of boson operators (10).
Because these one-body operators commute with the norm operator, the
interaction will as well.  Normal ordering we get
\begin{equation}
\hat{V}_B  = \sum_{\mu \nu} \left \langle \mu  \right |
V \left | \nu \right \rangle \left[
{b}^\dagger_\mu {b}_\nu
+ 2 \sum_{\sigma \sigma^\prime} \sum_{\tau \tau^\prime}
{\rm tr \,}
\left (
{\bf A}_\sigma {\bf A}^\dagger_\mu
{\bf A}_{\sigma^\prime} {\bf A}^\dagger_\tau
{\bf A}_\nu {\bf A}^\dagger_{\tau^\prime}
\right )
{b}^\dagger_\sigma {b}^\dagger_{\sigma^\prime}
{b}_\tau {b}_{\tau^\prime}\right]
\end{equation}
and in general one can find a image Hamiltonian
$ \hat{H}_B = \hat{T}_B + \hat{V}_B$.
This result, and its relation to other mappings such as the
non-Hermitian Dyson mapping, is found in Marshalek \cite{marshalek}.

Thus any boson representation of a Hamiltonian factorizes:
$\hat{\cal H}_B = \hat{\cal N}_B \hat{H}_B$.  Since the norm operator is a
function of the ${\rm SU}(2\Omega)$ Casimir operators it commutes with the
boson
images of fermion operators \cite{JG}, and one can simultaneously
diagonalize both $\hat{H}_B$ and $\hat{\cal N}_B$.  Then
Eqn.~(\ref{BosonEigen1}) becomes
\begin{equation}
 \hat{H}_B\left | \Phi_p \right ) =
E^\prime_p \left | \Phi_p \right ).
\label{EigenBoson}
\end{equation}
where $E^\prime_p = E_p$ for the physical states, but $E^\prime_p$ for the
spurious states is no longer necessarily zero.
The boson Hamiltonian $\hat{H}_B$ is by construction Hermitian and, if one
starts with at most only two-body interactions between fermions, has at most
two-body
boson interactions.  All physical eigenstates of the original fermion
Hamiltonian will have counterparts
in (\ref{EigenBoson}).  It should  be clear that transition amplitudes
between physical eigenstates will be preserved. Spurious states will also
exist but, since the norm operator $\hat{\cal N}_B$ commutes with the boson
image
Hamiltonian $\hat{H}_B$, the physical eigenstates and the spurious states
will not admix. 
Identification of the spurious states is a serious though tractable
problem, as $\hat{\cal N}_B$ annihilates such states; furthermore
spurious states can be shifted up in the spectrum through the use of the
Park operator\cite{Klein}, $\hat{M} =  \colon \hat{C}_2 \colon
+4 N(N-1) $, which
has zero eigenvalue for physical states and a positive definite spectrum for
spurious states.

A more critical question however is that of
truncating the boson Fock space, by which we mean
using states constructed from a restricted
set of fermion pairs/bosons denoted
by $\left \{ \bar{\alpha} \right \}$;  the
operators in this space we denote by $\left [ \hat{\cal N}_B \right ]_T$,
$\left [ \hat{\cal H}_B \right ]_T$, and so on, and are straightforward
to construct:  for example, the norm operator is
\begin{equation}
\left [ \hat{\cal N}_B \right ]_T
=
\colon \exp \left (  \sum_{k=2}^\infty
{ { (-2)^{k-1}} \over k }  {\rm tr \,}
\left [ {\bf P} \right ]^k_T \right ) \colon
\end{equation}
where
$\left [ {\bf P} \right ]^k_T = \sum_{\bar{\sigma}, \bar{\tau}}
b^\dagger_{\bar{\sigma}} b_{\bar{\tau}}
{\bf A}_{\bar{\sigma}} {\bf A}^\dagger_{\bar{\tau}}$.
These truncated
representations still exactly preserve the matrix elements in the
restricted fermion space:
$ \left ( \phi_{\bar{\alpha}} \right |
\left [ \hat{\cal N}_B\right ]_T \left | \phi_{\bar{\beta}} \right )
= \left \langle \psi_{\bar{\alpha}} \right. \left |
\psi_{\bar{\beta}} \right \rangle$ and so on.
 This is true even when the truncated set of
fermion pairs represented do not form a closed subalgebra, a fact apparently
overlooked previously \cite{doba}.  It should be evident that the
truncated representations still do not mix physical and spurious states.

Although the representations remain exact under truncation,
the factorization  into the image does not persist in general:
 $\left [ \hat{\cal H}_B \right ]_T
\neq   \left [ \hat{H}_B \right ]_T \left [ \hat{\cal N}_B \right ]_T$.
This was recognized by Marshalek \cite{marshalek} and arises because
the completeness relation (\ref{completeness}) in general
is only satisfied if the  complete Fock space is used.
(An alternate formulation found in \cite{marshalek} does not require the
complete Fock space, but mixes physical and spurious states and so always
requires a projection operator.)

We can however  find
a sufficient condition such that a factorization
\begin{equation}
\left [ \hat{\cal H}_B \right ]_T
=  \bar{H}_B \left [ \hat{\cal N}_B \right ]_T
\label{NewFactor}
\end{equation}
{\it does} exist, with
$  \bar{H}_B$ at most two-body;  then with a further, stricter condition can
guarantee $\bar{H}_B$ is Hermitian and commutes with
$ \left [ \hat{\cal N}_B \right ]_T$.
First,
consider a partition of the single fermion states labeled by $i = (i_a, i_c)$,
where the dimension of each subspace is $2\Omega_a$, $2\Omega_c$  so that
$\Omega = 2 \Omega_a \Omega_c$.  We denote the amplitudes for the truncated
space as ${\bf A}^{\dagger}_{\bar{\alpha}}$ and assume they can be factored,
$ ({\bf A}^{\dagger}_{\bar{\alpha}})_{ij} = ({\bf K}^{\dagger})_{i_a j_a}
\otimes (\bar{{\bf A}}^{\dagger}_{\bar{\alpha}})_{i_c j_c}$, with
${\bf K}^{\dagger} {\bf K} = {\bf K K}^{\dagger} = \frac{1}{2 \Omega_a}$
${{\bf K}^T} = (-1)^p {\bf K}$,
where $p = 0$ (symmetric) or $p = 1$ (antisymmetric).

Furthermore we assume the completeness relation (\ref{completeness}),
which was crucial for
proving that $\hat{\cal H}_B = \hat{H}_B \hat{\cal N}_B$, is valid for the
truncated space,
\begin{equation}
\sum_{\bar{\alpha}} (\bar{\bf A}^{\dagger}_{\bar{\alpha}})_{i_c j_c}
(\bar{\bf A}_{\bar{\alpha}})_{j_c^{\prime} i_c^\prime}
= \frac{1}{2} \left[
\delta_{i_c, i^{\prime}_c} \delta_{j_c ,j^{\prime}_c} - (-1)^p \delta_{i_c,
j^{\prime}_c} \delta_{i^{\prime}_c,j_c} \right].
\label{newcomplete}
\end{equation}
A necessary condition is that the the set of operators,
$\left \{ \hat{A}_{\bar{\alpha}}, \hat{A}_{\bar{\beta}}^\dagger,
\left [ \hat{A}_{\bar{\alpha}}, \hat{A}_{\bar{\beta}}^\dagger \right ]
\right \}$ form a closed subalgebra, as in the example given below.

The norm operator in the truncated space then becomes
\begin{equation}
\left[\hat{\cal{N}}_B\right]_T  = \colon {\rm exp} \sum_{k = 2}
\left(\frac{-1}{\Omega_a}\right)^{k-1} \frac{1}{k}
{\rm tr} (\bar{\bf P}^k) \colon ,
\end{equation}
where $\bar{\bf P} = \sum_{\bar{\sigma} \bar{\tau}} b^{\dagger}_{\bar{\sigma}}
b_{\bar{\tau}} \bar{\bf A}_{\bar{\sigma}}
\bar{\bf A}^{\dagger}_{\bar{\tau}}$ so that
 $\left[{ \bf P}\right]_T = \left(\frac{1}{2
\Omega_{a}}\right)\bar{\bf P}$.
In this case the boson image of a one-body operator is the truncation of the
boson image in the full space,
\begin{equation}
\left [ \hat{\cal T}_B \right ]_T =
\left [ \hat{ T}_B \right ]_T
\left [ \hat{\cal N}_B \right ]_T ,
\end{equation}
\begin{equation}
\left [ \hat{ T}_B \right ]_T =
2\sum_{ \bar{\sigma}, \bar{\tau}}
{\rm tr \,} \left ( {\bf A}_{\bar{\sigma}} {\bf T}
{\bf A}^\dagger_{\bar{\tau}} \right )
b^\dagger_{\bar{\sigma}} b_{\bar{\tau}}.
\end{equation}

The representation of a two-body
interaction can be factored into a boson image times the
truncated norm,
\begin{equation}
\left [ \hat{\cal V}_B \right ]_T =
 { \bar{V}}_B
\left [ \hat{\cal N}_B \right ]_T ;
\end{equation}
but in the case where  only (\ref{newcomplete}) holds, however,
$\bar{V}_B$, while finite (1+2-body), is not simply related to
$\left [ V_B \right ]_T$ as is the case for one-body operators
and in fact is not
necessarily Hermitian.  We will discuss this general case elsewhere
\cite{JG}.  Suppose one has the additional condition
\begin{eqnarray}
\sum_{\mu, \nu} \left \langle \mu \left | V \right | \nu \right \rangle
\sum_{i_a, j_a}
\left (  {\bf A}_\nu \right )_{ i_a i_c, j_a j_c}
\left ( {\bf A}^\dagger_\mu  \right )_{j_a j_c^\prime, i_a i_c^\prime}
\nonumber \\
= N_a
\sum_{\mu, \nu}
 \left \langle \mu \left | V \right | \nu \right \rangle
\sum_{i_a, j_a}
\left (  {\bf A}_\nu \right )_{ i_a i_c, j_a j_c}
\left ( {\bf K}^\dagger \right ) _{ j_a, i_a}
\sum_{i_a^\prime, j_a^\prime}
\left ( {\bf K} \right ) _{ i_a^\prime, j_a^\prime}
\left ( {\bf A}^\dagger_\mu  \right )_{j_a^\prime j_c^\prime,
i_a^\prime i_c^\prime}
\label{VFactorCondition}
\end{eqnarray}
where the factor $N_a=\Omega_a(2\Omega_a+(-1)^p) $ is the number of pairs in
the excluded subspace; while condition(\ref{VFactorCondition}) looks
complicated there are interactions that satisfy it, for example,
two-body interactions constructed from one-body operators
$\hat{V} = \hat{T}_{\bar{\alpha} \bar{\beta}}
\hat{T}_{\bar{\alpha}^\prime \bar{\beta}^\prime}$ where
$\hat{T}_{\bar{\alpha} \bar{\beta}} = \left [
A^\dagger_{\bar{\alpha}} , A_{\bar{\beta}} \right ]$. When
(\ref{VFactorCondition}) is satisfied
then $\bar{V}_B$ is Hermitian and although
 $\bar{V}_B \neq \left [ V_B \right ]_T$ they are simply related:
\begin{equation}
\bar{V}_B =
\sum_{\bar{\sigma},\bar{\tau}}
\left \langle \bar{\sigma} \left | V \right | \bar{\tau} \right \rangle
b^\dagger_{\bar{\sigma}} b_{\bar{\tau}}
+
 2 f_{\Omega_a}
\sum_{{\mu},{\nu}}
\left \langle {\mu} \left | V \right | {\nu} \right \rangle
\sum_{  \bar{\sigma} \bar{\sigma}^\prime,
\bar{\tau} \bar{\tau}^\prime }
{\rm tr \,}
\left (
{\bf A}_{\bar{\sigma}} {\bf A}^\dagger_{{\mu}}
{\bf A}_{\bar{\sigma}^\prime} {\bf A}^\dagger_{\bar{\tau}}
{\bf A}_{{\nu}} {\bf A}^\dagger_{\bar{\tau}^\prime}
\right )
b^\dagger_{\bar{\sigma}} b^\dagger_{\bar{\sigma}^\prime}
b_{\bar{\tau}} b_{\bar{\tau}^\prime}
\label{SimpleV}
\end{equation}
with  $f_{\Omega_a} = 4\Omega_a^2/N_a $
renormalizing the two-boson part of $ \left [ V_B \right ]_T$
by a factor  which
ranges from unity (full space) to 2 for a very small subspace.

The SO(8) and Sp(6) models [15] belong to a class of models which have a
subspace for which (\ref{newcomplete}) is valid.
In these models the shell model orbitals have
a definite angular momentum $\vec{{\bf j}}$ and
are partitioned into a pseudo
orbital angular momentum $\vec{{\bf k}}$ and pseudospin $\vec{{\bf
 i}}$,
$\vec{\bf j}
= \vec{{\bf k}} + \vec{{\bf i}}$.  The amplitudes are then given
as products of Clebsch-Gordon coefficients,
$\left(A^{\dagger}_{\alpha}\right)_{ij} = \frac{(1 + (-1)^{K + I})}{2}
(k \, m_i, k\, m_j |K_\alpha \, M_\alpha) \,
(i \, \mu_i, i\, \mu_j |I_\alpha \, \mu_\alpha)$,
where $K$ and $I$ are the total pseudo orbital angular momentum and pseudospin
respectively of the pair of nucleons.
For the SO(8) model  ${\bf i} =
\frac{3}{2}$ and one  considers
the subspace of pairs with $K = 0$ $(p = 0)$,
$(\bar{A}^{\dagger}_{\bar{\alpha}})_{ij} = \frac{(1 + (-1)^I)}{2}
( i \, \mu_i, i \, \mu_j | I_\alpha \, \mu_\alpha)$;
in the Sp(6) model  ${\bf k} = 1$ and one considers
 the subspace with $I = 0$ $(p = 1)$,
$(\bar{A}^{\dagger}_{\bar{\alpha}})_{ij} = \frac{(1 + (-1)^K)}{2}
(k \, m_i, k\, m_j | K_\alpha \, M_\alpha)$. For these subspaces the
completeness relation (\ref{newcomplete}) holds, and the
multipole-multipole interactions of \cite{Talmi2} satisfy
(\ref{VFactorCondition}) and
yield finite, Hermitian images as defined in (\ref{SimpleV}).

Not all interactions satisfy (\ref{VFactorCondition}); for example, the
pairing interaction never does \cite{JG} except in the full space.
In such cases the
factored boson image is usually non-Hermitian \cite{JG}.
Since the norm operator and the boson representation of Hermitian
fermion operators are Hermitian, and under factorization
 $\left [ \hat{\cal H}_B \right ]_T =
 \bar{H}_B\left [ \hat{\cal N}_B \right ]_T =
\left [ \hat{\cal H}_B \right ]_T^\dagger =
\left [ \hat{\cal N}_B \right ]_T \bar{H}_B^\dagger$,
the boson image in the truncated space is Hermitian if and only if it
commutes with the norm.
If the image commutes with the norm then they can be simultaneously
diagonalized and the image does not mix physical and spurious states.
If the image does not commute, however, as is the case for the
non-Hermitian Dyson mapping, then it can mix physical and spurious states.
This is because the norm is a singular operator with null eigenvalues;
even though the representations have zero matrix elements between physical
and spurious states, by factoring out the norm one essentially is dividing
zero by zero which can lead to mixing.

For truncation schemes where the conditions (\ref{newcomplete})
or  (\ref{VFactorCondition}) do not hold,
it is better to define a hermitian image Hamiltonian
\begin{equation}
\bar{H}_B \equiv
\left [ \tilde{\cal N}_B \right ]_T^{-1/2}
\left [ {\cal H}_B \right ]_T
\left [ \tilde{\cal N}_B \right ]_T^{-1/2} .
\label{NumericalTruncate}
\end{equation}
Because the norm is a singular operator
it cannot be inverted. Instead $\left [ \tilde{\cal N}_B \right ]_T^{-1/2}$
is calculated from the norm only in the physical subspace,
with the zero eigenvalues which annihilate the spurious states retained.
Then $\bar{H}_B$ does not mix physical and spurious states.
 When the image commutes with the norm this is
equivalent to our previous definition.

$\bar{H}_B$ is not necessarily finite; it may have an infinite expansion
which for purposes of discussion we sketch as
\begin{equation}
 \bar{H}_T = \theta_1 b^\dagger b + \theta_2  b^\dagger b^\dagger b b
+ \theta_3  b^\dagger b^\dagger b^\dagger b b b + \ldots
\label{HTruncate}
\end{equation}
We now argue that calculation of the coefficients $\theta_k$ for the $k$-body
term can be done in a straightforward manner.   Because the norm operator
is of the form (\ref{NormBosonRep}), the operator
$\left [\tilde{\cal N}_B \right ]_T^{-1/2}$ must be of a similar form.  Further
the $k$-body term can be calculated by diagonalizing
$\left [ {\cal N}_B \right ]_T$ in the truncated $k$-body Fock space, and
$\theta_k$ at most requires the $k$-body terms of
$\left [ \tilde{\cal N}_B \right ]_T^{-1/2}$ and
$\left [ {\cal H}_B \right ]_T$.   Then one can consider the convergence
of the series (\ref{HTruncate}) as a function of $k$. A rough estimate
is that, for an $N$-boson Fock space, one can truncate to the $\ell$-body
terms if for $k > \ell$,  $\theta_k$ is sufficiently small compared to
$\theta_\ell \times (N-k)!/(N-\ell)!$; the strictest condition is to
require $\theta_k \ll \theta_\ell /(k-\ell)!$.

   Under a finite termination of (\ref{HTruncate})
 $\bar{H}_B$ can again mix physical and spurious
states, but if the remainder is small the mixing may be small. (If the
truncation in the boson Fock space is severe enough the boson Fock space
may not even contain any spurious states, as for the SO(8) model below
half-filling.)

We emphasize the utility of this ``linked-cluster'' expansion similar to
that of \cite{Kishimoto}.   Suppose the series (\ref{HTruncate})
converges fast enough that truncation to 4-body terms is sufficient
even for states with 5 or 6 bosons by the criterion given above.  If one
takes just $J=0,2$ bosons (as for the interacting boson model), by
partitioning via angular momentum the largest matrix that needs to be
inverted to calculate the 4-body terms is of dimension 4.
This approximate image Hamiltonian, although calculated in a 4-boson system,
would be applicable for $N$-boson systems with $N \gg 4$, failing only
when a significant number of spurious states enter the Fock space.

As a final example, consider the pairing interaction in the seniority-0
subspace, where there is only one state, $(S^\dagger)^N \left | 0 \right
\rangle$, with
$S^\dagger \equiv (4\Omega)^{-1/2} \sum_i \epsilon_i
a^\dagger_i a^\dagger_{\bar{\imath}}$; here $\bar{\imath}$ is the
time-reverse of the state $i$, $\epsilon_i = \pm 1$ and
$\epsilon_{\bar{\imath}} = - \epsilon_i$.   Under
naive truncation the pairing interaction,
$\hat{V}^P = -GS^\dagger S$, gets mapped to
\begin{equation}
\left [\hat{V}^P_B \right ]_T = -G\left( s^\dagger s
+ {1 \over 2\Omega^2} s^\dagger s^\dagger ss \right )
\end{equation}
which does not give the correct ground state energy.
Because condition (\ref{VFactorCondition}) does not hold one cannot use
(\ref{SimpleV}) to obtain the image; instead (\ref{NumericalTruncate}) yields
the correct answer:
\begin{equation}
\bar{V}^P_B  = -G\left( s^\dagger s
- {1 \over \Omega } s^\dagger s^\dagger ss\right) .
\end{equation}
Applications to realistic systems will be discussed in future investigations.

This work was supported by the U.~S.~Department of Energy.

\pagebreak

\end{document}